# DeepFixel: Crossing white matter fiber identification through spherical convolutional neural networks

Adam M. Saunders*[a], Lucas W. Remedios[b], Elyssa M. McMaster[a], Jongyeon Yoon[b], Gaurav Rudravaram[a], Adam Sadriddinov[b], Praitayini Kanakaraj[a], Bennett A. Landman[a-e], Adam W. Anderson[c-e]

[a]Department of Electrical and Computer Engineering, Vanderbilt University, Nashville, TN, USA; [b]Department of Computer Science, Vanderbilt University, Nashville, TN, USA; [c]Vanderbilt University Institute of Imaging Science, Vanderbilt University Medical Center, Nashville, TN, USA; [d]Department of Radiology and Radiological Sciences, Vanderbilt University Medical Center, Nashville, TN, USA;  [e]Department of Biomedical Engineering, Vanderbilt University, Nashville, TN, USA

## ABSTRACT

Diffusion-weighted magnetic resonance imaging allows for reconstruction of models for structural connectivity in the brain, such as fiber orientation distribution functions (ODFs) that describe the distribution, direction, and volume of white matter fiber bundles in a voxel. Crossing white matter fibers in voxels complicate analysis and can lead to errors in downstream tasks like tractography. We introduce one option for separating fiber ODFs by performing a nonlinear optimization to fit ODFs to the given data and penalizing terms that are not symmetric about the axis of the fiber. However, this optimization is non-convex and computationally infeasible across an entire image (approximately $1.01 \times 10^6$ ms per voxel). We introduce DeepFixel, a spherical convolutional neural network approximation for this nonlinear optimization. We model the probability distribution of fibers as a spherical mesh with higher angular resolution than a truncated spherical harmonic representation. To validate DeepFixel, we compare to the nonlinear optimization and a fixel-based separation algorithm of two-fiber and three-fiber ODFs. The median angular correlation coefficient is 1 (interquartile range of 0.00) using the nonlinear optimization algorithm, 0.988 (0.317) using a fiber bundle elements or "fixel"-based separation algorithm, and 0.973 (0.004) using DeepFixel. DeepFixel is more computationally efficient than the non-convex optimization (0.32 ms per voxel). DeepFixel's spherical mesh representation is successful at disentangling at smaller angular separations and smaller volume fractions than the fixel-based separation algorithm.

## 1. INTRODUCTION

Diffusion magnetic resonance imaging (MRI) is an imaging modality that is sensitive to the random thermal motion of water molecules. The diffusion motion of water molecules is highly restricted in the white matter fibers of the brain. Diffusion MRI provides measurements from which we can reconstruct information about these white matter fibers and therefore the structural connectivity of the brain. After forming local models for the orientation of white matter fibers in a voxel from the diffusion MRI, we can follow pathways of high coherence between voxels to estimate tracts that describe connections between regions of the brain.[1]

White matter fibers are several orders of magnitude smaller than the resolution of diffusion MRI scans, leading to the crossing fiber problem, where there are multiple orientations of fibers within a single voxel (Fig. 1). The geometry of fibers within a voxel is highly complex, complicating the estimation of local fiber orientation. Crossing fibers occur in approximately 60-90% of white matter voxels.[2] Crossing fibers are not simply a result of limited spatial resolution, and they appear in ex vivo diffusion MRI and histology at higher resolutions than implemented for in vivo imaging.[3] Other complex geometries like fanning or bending of fibers, as well as the convergence of fibers into a single voxel bottleneck, can complicate analysis and lead to partial volume effects where multiple tissue types occupy a voxel[4]. One method for disentangling the crossing fibers is fiber bundle elements or "fixels." Fixels model the fiber bundles as directions weighted by the volume fraction occupied by the fiber bundle.[5] Fixels can provide a framework for mapping parameters of interest for cross-subject comparison.[6]

*adam.m.saunders@vanderbilt.edu

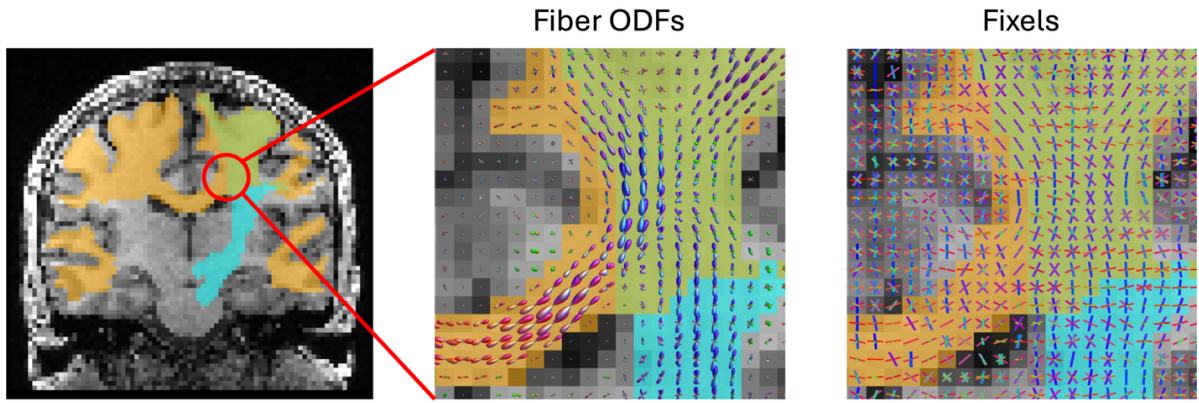

Figure 1. Fiber orientation distribution functions (ODFs) near the intersection of the corpus callosum (orange) and corticospinal (blue) tracts show multiple directions with a high diffusion signal, a sign of the crossing fiber problem. Fitting vectors to the signal using fod2fixel from the mrtrix3 package,[22] we can see that there are multiple fibers in each voxel (crossing lines at the intersection of the bundles on the right). This geometry complicates analysis and downstream tasks like tractography. The images are generated from data publicly available from Rokem[35] with tract segmentations generated from TractSeg.[36]

There are many models for estimating local fiber orientation and describing diffusion in the brain. The "Standard Model" of diffusion models three compartments: a distribution of sticks and ellipsoids representing intra-axonal and extra-axonal diffusion and an isotropic ball representing free water.[7] We can describe many parametric models for fiber orientation and diffusion within this framework by adding constraints or removing compartments. Single-tensor models fail to account for crossing fibers, but extensions to multiple tensors can describe multiple fibers.[8,9] Multi-compartment models such as neurite orientation dispersion and density imaging (NODDI[10]) and the ball-and-stick model[11] model the diffusion signal as a combination of responses from several tissues. An alternative to models derived from the Standard Model are model-free descriptions of fiber orientation. Spherical deconvolution methods model the fiber orientation distribution function (ODF) as a function on the unit sphere convolved with the diffusion signal response to a single fiber. This response can be estimated from the data as in constrained spherical deconvolution,[12] or it can be modeled from a continuous distribution of tensors.[13] Multi-compartment models often require diffusion MRI acquired with multiple b-values or shells, while spherical deconvolution only requires one shell, though we can use multi-shell data to estimate multiple fiber ODFs of different tissue types.[14]

Recent work in deep learning has allowed for more efficient and better estimations of fiber ODFs. For example, to take advantage of the geometric structure of fiber ODFs, one can form networks invariant to spatial rigid transforms and rotations on the hemisphere.[15] Other networks have focused on explaining fiber geometry. Convolutional neural networks can provide estimates of the number of fibers in each voxel[16] and their orientations.[17] Spherical deconvolution networks have been applied for estimation of fiber ODFs and simultaneous estimation of number of fibers.[18] Karimi et al. used a fully-connected network to estimate the angle from the nearest fiber element on a set of discrete directions to estimate fiber orientations.[19]

Typical fixel-based analysis uses weighted peak directions to represent fibers.[5] When there are multiple peaks in an observed distribution of fiber orientations, the actual fibers may be shifted relative to the peaks of the observed distributions. Using the peak directions of the distribution to represent the fibers is biased, while deconvolving the individual fiber distributions is computationally expensive. In this work, we describe a nonlinear non-convex optimization algorithm for separating single-fiber ODFs from multi-fiber ODFs. Then, we present a massively accelerated deep learning approximation for this method.

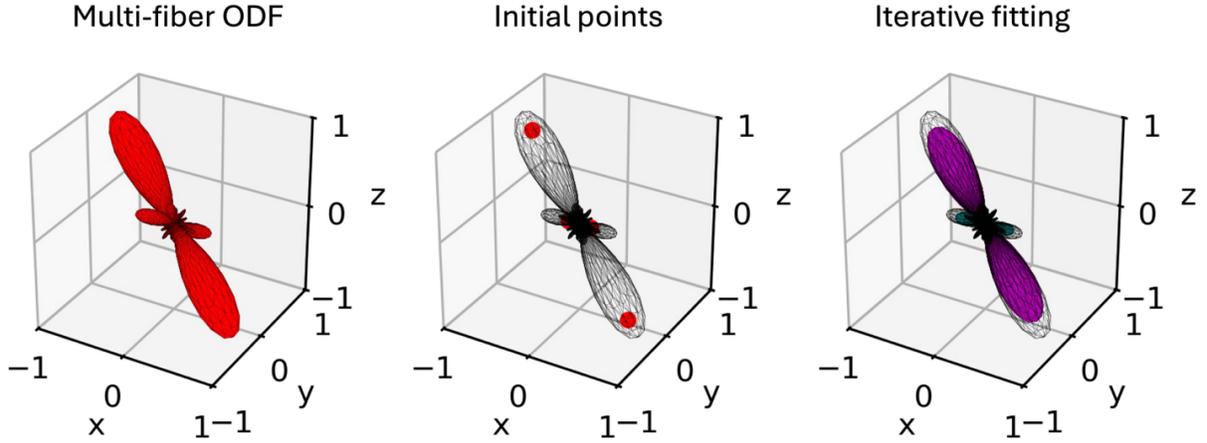

Figure 2. FISSILE first uses a watershed algorithm to determine initial points for the individual fibers' orientations. Then, FISSILE iteratively fits axially symmetric fiber ODFs to the data, a computationally intensive nonlinear non-convex search, taking approximately 17 minutes to resolve three crossing fibers on a computer with 32 Intel Zeon E5-2630 CPUs.

## 2. METHODS

**Simulation data**

To investigate the feasibility of using a deep learning algorithm to approximate an algorithm for separating crossing fiber ODFs, we use simulated data. We use simulated data because we can control the direction and size of the fiber ODF, and this data can serve as ground truth to test the sensitivity of fiber separation algorithms. A single-fiber ODF $Y_i(\theta, \phi)$ at a given spherical angle $(\theta_i, \phi_i)$ with volume fraction $v_i$ can be represented using a spherical harmonic basis with coefficients $Y_l^m$ of order $l$ and phase $m$. The fiber ODF represented by a single fiber is a hemispherical Dirac delta function that we truncate to order 6. To enforce antipodal symmetry, we only consider spherical harmonics of even order. We simulate crossing fibers by adding the coefficients of two or three single-fiber ODFs, selected using random directions sampled uniformly on the unit sphere. To sample the volume fractions, we use a symmetric Dirichlet distribution with a concentration parameter of $\alpha = 1$. For example, for three fibers:

$$p(v_1, v_2, v_3) = \frac{\Gamma(3\alpha)}{\Gamma(\alpha)^3} \prod_{i=1}^{3} v_i^{\alpha-1}, \tag{1}$$

where $v_1 + v_2 + v_3 = 1$ and $\Gamma(\cdot)$ is the gamma function.[20]

This distribution allows for uniform sampling of volume fractions that sum to one for both two and three fibers. We generated a set of 250 two-fiber and 80 three-fiber ODFs for validating the crossing fiber separation algorithms.

**Fod2fixel and FISSILE optimization algorithms for crossing fiber separation**

We apply two optimization algorithms for separating crossing fiber ODFs as baselines for comparison. First, we compare with the fixel-based ODF segmentation implemented in the fod2fixel command in the mrtrix3 package.[21,22] Fod2fixel first separates a multi-fiber ODF into lobes representing each individual fiber. To perform this task, fod2fixel samples the fiber ODF at 1281 directions on the unit sphere. Then, fod2fixel compares samples to their neighbors on a level set of the surface of a sphere with decreasing radius. If a sample's neighbors have a larger amplitude than the sample, the neighbor is assigned to a new lobe. After segmenting the lobes, the weighted mean direction of each lobe is a fixel.

Second, we use a nonlinear fitting algorithm that we introduce, called fiber segmentation by signal-splitting and iterative least squares estimation (FISSILE, meaning "easily split"). FISSILE is a nonlinear non-convex optimization algorithm that fits single-fiber ODFs to the total ODF (Fig. 2). First, given a multi-fiber ODF, FISSILE samples the function on the unit sphere on a grid of angles and uses a watershed algorithm to segment the ODF into its corresponding lobes. Then, FISSILE finds an initial starting frame for the search, combining neighboring lobes if their angular separation is less than 35° or peak magnitude is less than 0.2. Around these initial frames, FISSILE optimizes to fit axially symmetric ODFs (with respect to the axis of the fiber), that is, penalizing spherical harmonic coefficients with phase $m \neq 0$.

Let $Y$ be a total multi-fiber ODF with $F$ fibers represented in a spherical harmonic basis with order $l$ and phase $m$. The goal of FISSILE is to find the single-fiber ODFs $y^{(f)}$ for $f = 1, 2, \ldots, F$. If we choose to represent $y^{(f)}$ as a set of spherical harmonic coefficients along the axis of the fiber (that is, the polar angle (0,0) lies along the axis of the fiber), then the total multi-fiber ODF can be represented as

$$\hat{Y} = \sum_{f=1}^{F} R^{(f)} \hat{y}^{(f)}, \tag{2}$$

where $R^{(f)}$ is the rotation matrix that rotates $y^{(f)}$ from the fiber axis to the standard image axis. To find fitted spherical harmonic coefficients $\hat{y}^{(f)}$, we first find the least squares solution to the following set of equations for an initial set of rotations:

$$\begin{cases} Y = \sum_{f=1}^{F} R^{(f)} y^{(f)} \\ 0 = m y_l^{m(f)} \text{ for all } f, l \text{ and } m \neq 0. \end{cases} \tag{3}$$

Once we have the least squares solution $\hat{y}^{(f)}$, we can find the cost of this solution. The cost of the fit consists of a fit cost, shape cost, and sign cost. The total cost is

$$\mathcal{L} = \mathcal{L}_{\text{fit}} + \mathcal{L}_{\text{shape}} + \mathcal{L}_{\text{sign}} \tag{4}$$

where

$$\begin{aligned} \mathcal{L}_{\text{fit}} &= \|\hat{Y} - Y\|_2 \\ \mathcal{L}_{\text{shape}} &= \max_f [\max(\hat{y}^{(f)}(90°, 0°) - \hat{y}^{(f)}(60°, 0°), \hat{y}^{(f)}(60°, 0°) - \hat{y}^{(f)}(30°, 0°), 0)] \\ \mathcal{L}_{\text{sign}} &= \begin{cases} 1, & \text{if sgn}\left(\hat{y}_0^{0(f)}\right) = -1 \text{ for any } f \\ 0, & \text{otherwise} \end{cases} \end{aligned} \tag{5}$$

are the individual losses. The fit cost is the $\mathcal{L}_2$ loss between the fitted spherical harmonic coefficients $\hat{Y}_l^m$ and the spherical harmonic coefficients of the total multi-fiber ODF $Y_m^l$. The shape cost encourages the values of the ODF to decrease as the polar angle $\theta$ increases from 0° to 90°. We found that a finer discretization (e.g., sampling at 30° and 45°) was sensitive to ringing from the spherical harmonics due to the truncation to order 6. The sign cost encourages ODFs that integrate to positive values on the total surface of the sphere. If the total cost is greater than $10^{-5}$ after fitting, we find the estimated fiber ODF with the greatest axial asymmetry and refit with another estimated fiber at that angle. The optimization algorithm was MATLAB's MultiStart algorithm.[23] The search is non-convex because of the axially symmetric constraints and therefore computationally intensive.

**Deep learning network for crossing fiber separation**

To disentangle the crossing fibers and approximate the FISSILE algorithm, we introduce DeepFixel, a deep learning network that estimates the probability of a fiber at a given spherical angle (Fig. 3). The input to the network is a multi-fiber ODF, and the output of the network is a probability distribution representing the distribution of fibers. We represent the input and outputs as hemispherical meshes sampled on a hierarchical equal area isolatitude pixelization (HEALPix) of 384 directions.[24] We choose to represent that inputs and outputs as meshes instead of the more compact spherical harmonic representation so that the network does not have to implicitly learn a conversion from spherical harmonics to mesh amplitudes to construct the probability distribution of fibers. The output is a spherical probability distribution representing the probability of finding a fiber along a given angle. While a fiber ODF can be considered as a probability distribution, the typical sparse representation of ODFs as a truncated set of spherical harmonic coefficients means that the angular resolution is limited by the order of the spherical harmonics selected. A maximum order of $l_{max} = 8$ struggles to resolve fibers smaller than an angular separation of about 40°, but higher orders are sensitive to noise in the diffusion MRI acquisition.[25]

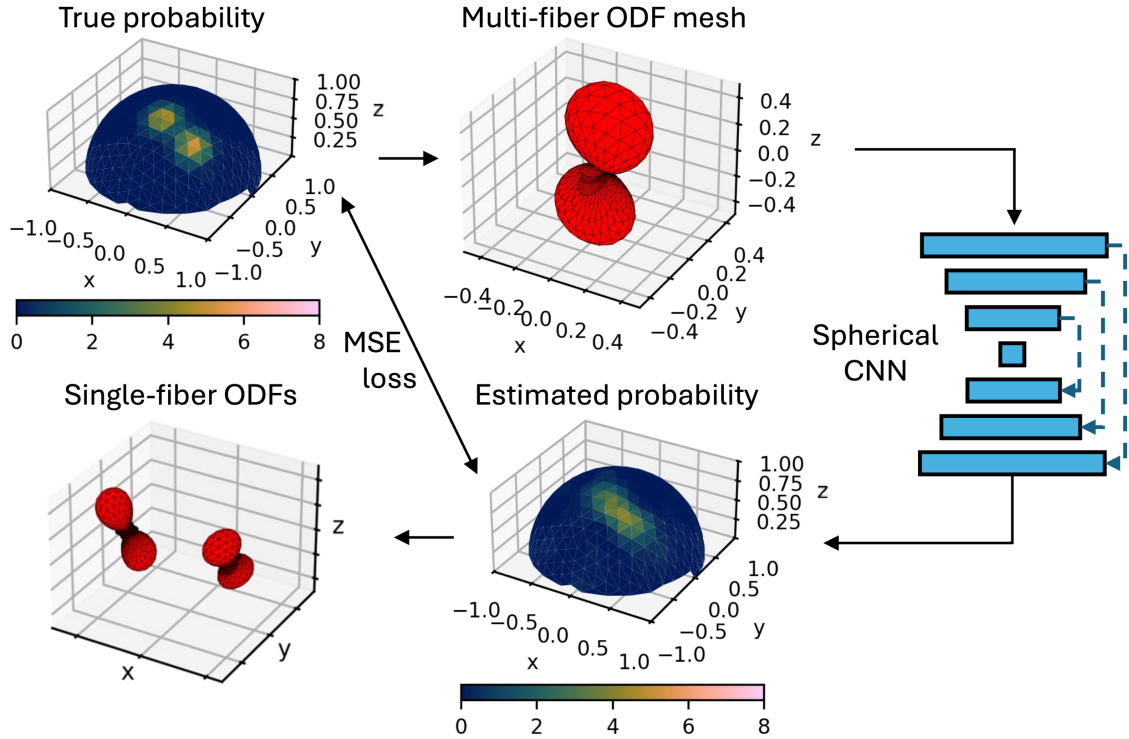

Figure 3. Our DeepFixel network models fiber distributions using a von Mises-Fisher distribution. As input, we use a multi-fiber ODF mesh, which captures the symmetric nature of orientation distribution functions. To account for rotational equivariance inherent in ODFs, we employ a spherical convolutional neural network (spherical CNN) following a U-Net architecture with skip connections between encoder and decoder layers. The network outputs an estimated probability distribution, which is compared to the ground truth using mean squared error loss. From the predicted distribution, we fit single-fiber ODFs based on peak directions and amplitudes to approximate FISSILE.

To address this limitation, we model the probability as a mixture of von Mises-Fisher distributions weighted by volume fraction. The von Mises-Fisher distribution is a distribution on the unit sphere that models a mean direction with a concentration around the mean, a spherical analogue of the Gaussian distribution.[26] We set the concentration parameter to $\kappa = 100$. The concentration parameter $\kappa$ determines the spread of the spherical von Mises-Fisher distribution about the mean direction. We chose $\kappa = 100$ because the full-width half-max (FWHM) of a von Mises-Fisher distribution with this parameter is approximately 13.5°, and the FWHM of an ideal fiber response of order 8 is approximately 11°.[13] From the output probability distribution, we can estimate the peak directions and amplitudes from the mesh to find the corresponding fiber orientations and volume fractions. To generate single-fiber orientations and volume fractions from DeepFixel's probability mesh, we find the peaks on the von Mises-Fisher distribution using Dipy.[27] We normalize the maxima's amplitudes by their sum to form the volume fraction.

We test two network architectures. The first is a standard multilayer perceptron (MLP) with 6 layers. The second is a spherical convolutional neural network (CNN) that ensures the network is equivariant to rotations. The spherical CNN has a UNet-style architecture with 4 downsampling layers and 5 upsampling layers.[15] We train the network on randomly-generated simulated multi-fiber ODFs with two and three single-fiber ODFs. We use the Adam optimizer,[28] a learning rate of $10^{-3}$ and a batch size of 512 voxels. The loss is mean squared error loss between the estimated probability distribution and the true probability distribution sampled on the same spherical mesh as the input to the network. The validation set consists of 512,000 simulated fiber ODFs, with the random number generator seeded so the validation set is consistent across batches. We train the network for 5,120,000 fiber ODFs, validating every 50 batches and stopping early if the validation loss had not decreased beyond the minimum validation loss in the last 250 batches.

To generate the single-fiber ODFs for fod2fixel and from the peak directions and volume fractions estimated from the DeepFixel output mesh, we model a hemispherical Dirac delta function at the given orientation. We truncate this function to spherical harmonics of maximum order 6 and multiply this function by the volume fraction to form the single-fiber ODF. We ran fod2fixel using 20 Intel Xeon W-2255 CPUs and DeepFixel using the same workstation with an Nvidia Quadro RTX 5000. Because of the computational burden of running FISSILE, we used a separate server with 32 Intel Zeon E5-2630 CPUs to run FISSILE. We evaluate the segmented fiber ODFs from FISSILE, fod2fixel, and DeepFixel using angular correlation coefficient (ACC), which measures the correlation between functions defined on the unit sphere.[13] ACC has been used in previous studies to compare two fiber ODFs.[13,29,30] An ACC of 1 is desirable because it indicates that the true fiber ODF and estimated fiber ODF are the same up to a normalization factor. To compare the peak directions and volume fractions generated by each method, we also report angular error and RMSE for volume fraction.

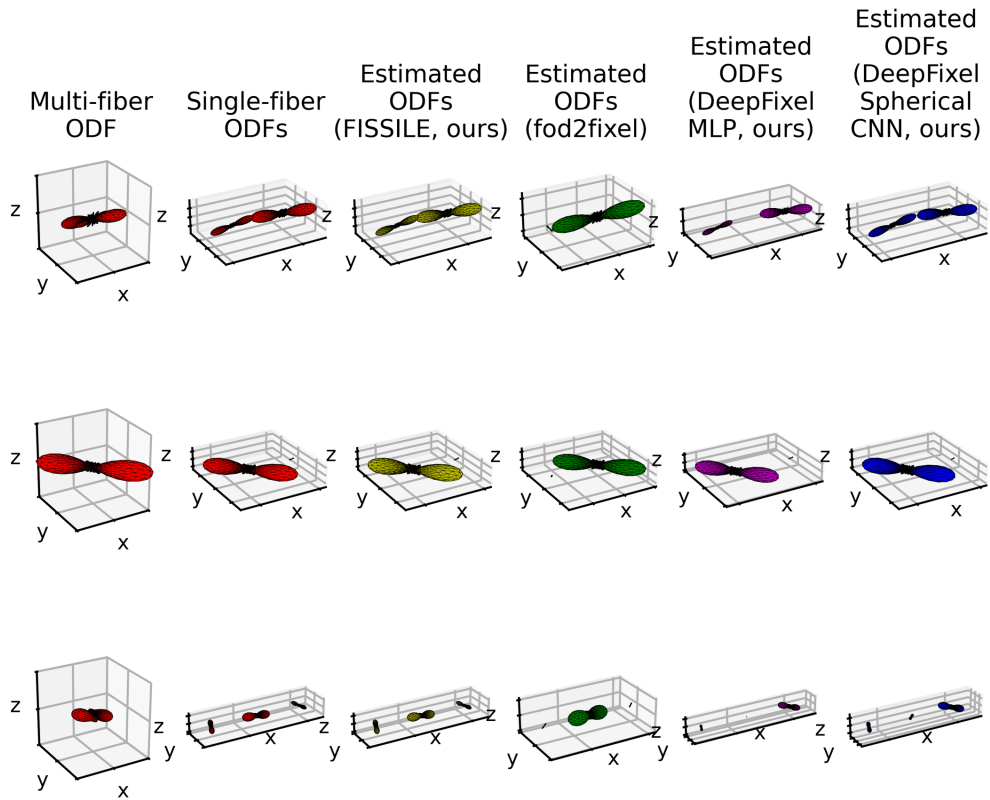

Figure 4. The single-fiber ODFs (separated along the x-axis) estimated from FISSILE are very accurate compared to the true single-fiber ODFs. The ODFs from fod2fixel are comparable to the estimated ODF's using the spherical CNN and the MLP. However, the ODFs estimated from fod2fixel are widely varying in terms of volume fraction and DeepFixel fails when volume fractions are small. (Top: 90th percentile, middle: 50th percentile, bottom: 10th percentile of average ACC across fibers for DeepFixel spherical CNN).

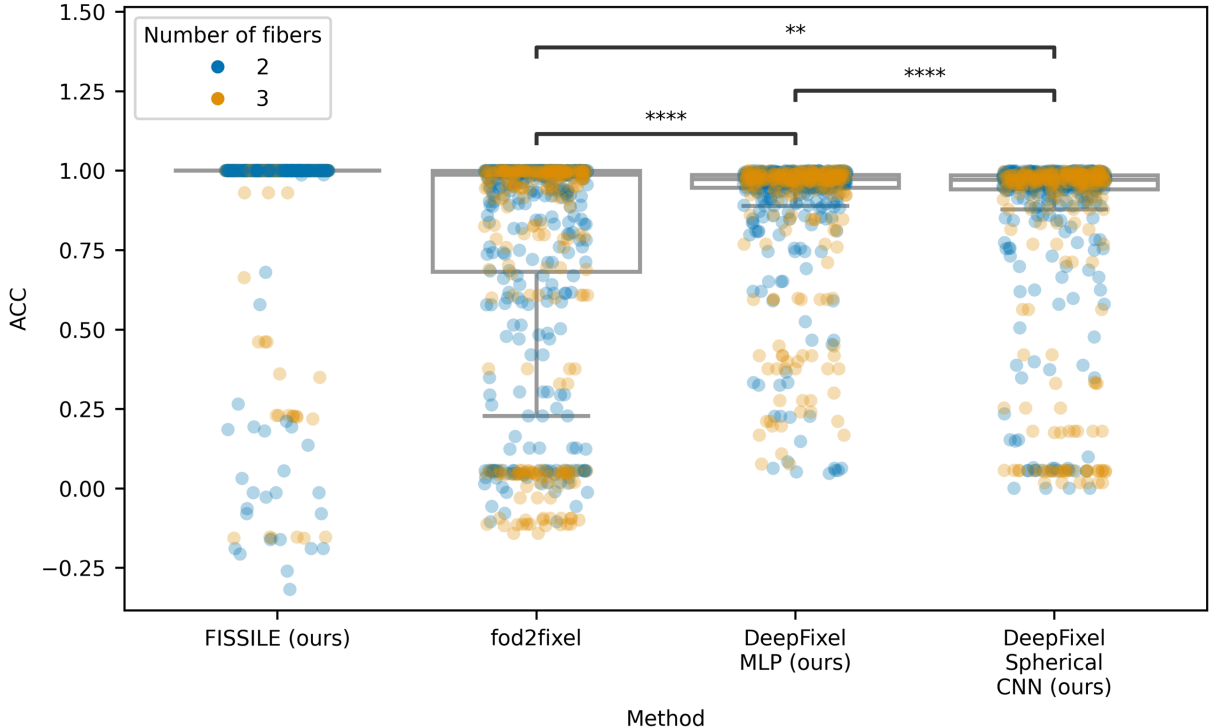

Figure 5. FISSILE computationally expensive search achieves an ACC of 1 for nearly all fibers. DeepFixel MLP and DeepFixel spherical CNN are significantly different from fod2fixel (**: $p < 0.01$, ****: $p < 0.0001$, Wilcoxon signed-rank test). Errors cluster around 0.1 for fod2fixel and DeepFixel spherical CNN when the peak finding fits to ringing due to the truncated spherical harmonic representation.

## 3. RESULTS

The single-fiber ODFs estimated from FISSILE are similar to the true single-fiber ODFs, even at very small volume fractions (Fig. 4). The ODFs from fod2fixel are close in direction to the original ODFs, but at widely varying volume fractions. DeepFixel's outputs for both the MLP and spherical CNN appear close to the original ODFs, though the network is sensitive at small volume fractions. The ACC of the ODFs from FISSILE is highly concentrated near 1, while fod2fixel and DeepFixel are more spread out (Fig. 5). The ACCs from DeepFixel MLP and DeepFixel spherical CNN are significantly different from fod2fixel ($p < 0.05$, Wilcoxon signed-rank test), and the effect size is moderate (Cohen's $d$ lies between 0.2 and 0.5 for both). While DeepFixel MLP and DeepFixel spherical CNN have significant differences in ACC, the effect size is small (Cohen's $d = 0.17$). Both DeepFixel networks produce ODFs whose ACC is more concentrated around the median than fod2fixel. All methods appear to perform similarly with both two and three fibers. We also tried increasing the resolution of the spherical mesh used to sample the probability distributions and varying the concentration parameter $\kappa$ for the von Mises-Fisher distribution, and we found no major effect on the network's performance. Table 1 shows the angular error and volume fraction error for each method.

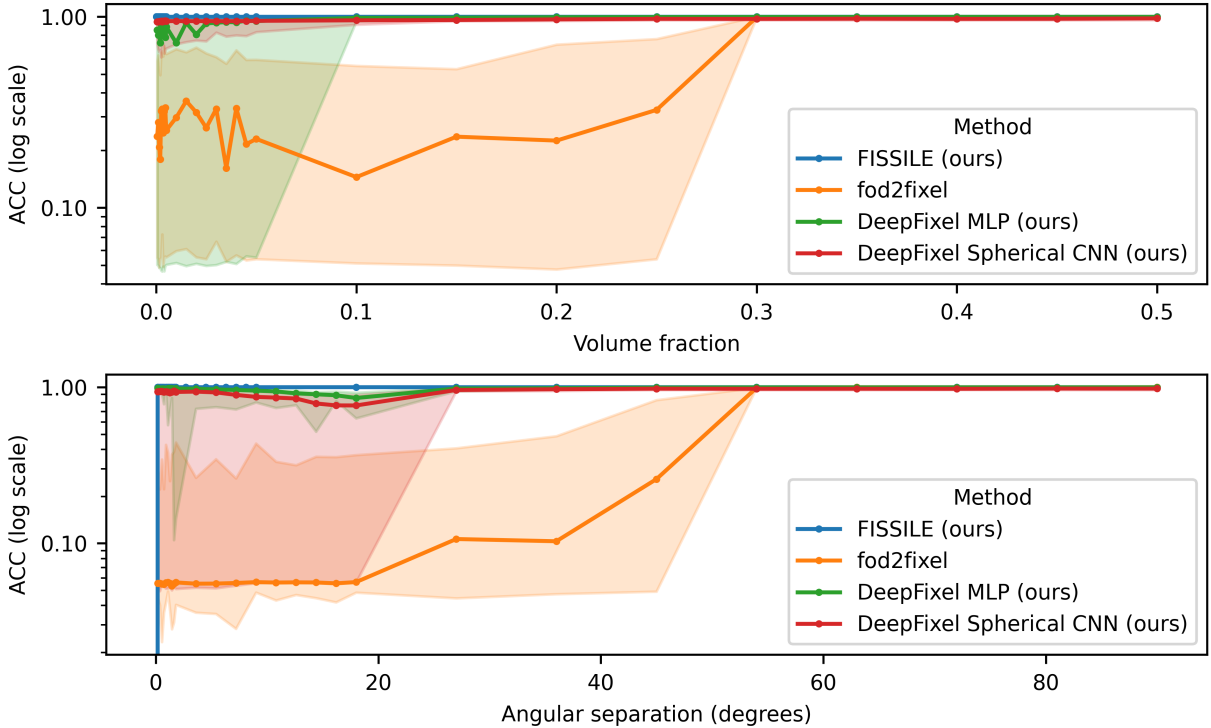

Figure 6. FISSILE successfully disentangles fibers across a wide range of volume fractions and angular separation. Fod2fixel fails to disentangle fibers at a volume fraction of approximately 0.25 and an angular separation of about 40°, while DeepFixel MLP and DeepFixel spherical CNN are less sensitive and disentangle fibers at a volume fraction of 0.1 and an angular separation of 25°. DeepFixel is less sensitive to volume fraction and angular separation than fod2fixel as an approximation for FISSILE. (Lines indicate median minimum ACC of two fibers, and bands indicate inter-quartile range.)

Table 1. Median (interquartile range) angular error and volume fraction error for each method. While FISSILE is extremely accurate, the computation time is impractical for large amounts of voxels. *: $p < 0.05$ compared to fod2fixel, †: $p < 0.05$ compared to DeepFixel MLP, ‡: $p < 0.05$ compared to DeepFixel spherical CNN with Mann-Whitney $U$-test.

| Method | Angular error (degrees) | Volume fraction error |
|---|---|---|
| FISSILE (ours) | $8.5 \times 10^{-7} (3 \times 10^{-6})^{*\dagger\ddagger}$ | $0.019 (0.023)^{*\dagger\ddagger}$ |
| fod2fixel | $24.8 (171.6)^{\dagger\ddagger}$ | $0.03 (0.06)^{\dagger\ddagger}$ |
| DeepFixel MLP (ours) | $3.7 (3.3)^{*\ddagger}$ | $0.15 (0.16)^{*\ddagger}$ |
| DeepFixel spherical CNN (ours) | $3.9 (9.7)^{*\dagger}$ | $0.05 (0.07)^{*\dagger}$ |

We investigated the sensitivity of these methods to volume fraction and angular separation. FISSILE performed well across nearly all values tested (Fig. 6). DeepFixel with both networks is less sensitive to volume fraction and angular separation than fod2fixel. In particular, fod2fixel began to fail at an angular separation of about 40°, consistent with previous literature in disentangling crossing fibers from ODFs represented as spherical harmonics.[25,31] DeepFixel separates fibers beyond this limit, though with a widely varying range of ACCs beyond a volume fraction of 0.1 and an angular separation of 25°.

## 4. DISCUSSION

DeepFixel provides a deep learning approximation for the accurate but computationally expensive FISSILE algorithm for crossing fiber disentangling. Notably, FISSILE and DeepFixel succeed in disentangling crossing fibers at smaller volume fraction and smaller angular separations than fod2fixel, separating angles beyond the previously reported limit of approximately 40°. We matched the estimated ODFs to the true ODFs by ACC, but this process is sensitive to small volume fractions because ACC normalizes the ODFs by their integral. Fod2fixel uses a different scaling for fiber ODFs so that the integral of the single-fiber ODF (called apparent fiber density) is a varying parameter across the brain.[32] FISSILE performed the best in terms of both angular error and volume fraction. DeepFixel spherical CNN resulted in less angular error and about the same volume fraction error as fod2fixel. In terms of computation time, FISSILE required 3 hours and 48 minutes to converge on 10 three-fiber ODFs ($1.01 \times 10^6$ ms per voxel or about 32 years for 1 million voxels). Fod2fixel required 41.72 seconds to converge on 1 million three-fiber ODFs (0.042 ms per voxel), DeepFixel MLP required 371.90 seconds (0.37 ms per voxel), and DeepFixel spherical CNN required 321.90 seconds (0.32 ms per voxel). While DeepFixel is more computationally expensive than fod2fixel, both are vastly more computationally efficient than FISSILE. Most of the computational burden of DeepFixel is spent finding the peaks of the von Mises-Fisher distribution output by the model. Our comparison of computation time is limited by the heterogeneous CPU/GPU architectures chosen due to the long computation time needed to run FISSILE.

We chose to empirically set the concentration parameter of the von Mises-Fisher distribution instead of varying it with the spread of different fiber ODFs. Using a constant concentration, DeepFixel does not account for the spread in fiber orientations. We could consider alternate bases to spherical harmonics like spherical needlets that can better represent crossing fiber geometries, albeit with a reduced ability to represent fiber fanning.[33] Because of the similar performance with the MLP, future applications of DeepFixel should use the spherical CNN network as it enforces rotational equivariance. We considered a single fiber ODF for DeepFixel, which means that we could estimate crossing fibers from single-shell acquisitions as well as multi-shell within a single framework.

Our study is limited by our use of simulated data with infinite signal-to-noise ratio (SNR), so our results represent an idealized model for separation of crossing fibers. While we assumed infinite SNR, in vivo diffusion MRI has an SNR of about 30, which introduces variability into the measured angular directions.[34] Future work should simulate this noise in the diffusion image before fitting fiber ODFs. A characterization of the downstream utility of DeepFixel should apply the network to tractography and in vivo diffusion MRI.

## 5. CONCLUSION

We introduced FISSILE, a computationally expensive but extremely accurate nonlinear non-convex optimization for disentangling crossing fiber ODFs in the brain. As an approximation to a computationally expensive nonlinear non-convex optimization, we developed a deep learning model called DeepFixel. By learning to represent the fiber orientations as a spherical probability distribution on a mesh instead of a spherical harmonic representation, the network was successful at separating fibers at smaller angular separations and smaller volume fractions than fixel-based decomposition of fibers ODF of simulated data. The code and testing dataset are available online in open source (https://github.com/MASILab/spherical_deep_fixel).

## ACKNOWLEDGEMENTS

This work was supported by NIH 1R01EB017230, NIH 1R01NS136743-A1, NIH P50HD103537 VKC, Integrated Training in Engineering and Diabetes grant number T32 DK101003, as well as DoD grant HT94252410563. This work was conducted in part using the resources of the Advanced Computing Center for Research and Education at Vanderbilt University, Nashville, TN. The Vanderbilt Institute for Clinical and Translational Research (VICTR) is funded by the National Center for Advancing Translational Sciences (NCATS) Clinical Translational Science Award (CTSA) Program, Award Number 5UL1TR002243-03. The content is solely the responsibility of the authors and does not necessarily represent the official views of the NIH.

We used generative AI to create code segments based on task descriptions, as well as to debug, edit, and autocomplete code. Additionally, generative AI technologies have been employed to assist in structuring sentences and performing grammatical checks. The conceptualization, ideation, and all prompts provided to the AI originated entirely from the authors' creative and intellectual efforts. We take accountability for the review of all content generated by AI in this work.